\documentclass[pre,twocolumn,floatfix,showpacs,showkeys]{revtex4}
\usepackage{hyperref}
\usepackage{amssymb}
\usepackage{amsfonts}
\usepackage{graphicx}
\usepackage{aecompl}
\usepackage{ae}

\newcommand{\Eq}[1]{Eq.~(\ref{#1})}
\newcommand{\Fig}[1]{Fig.~\ref{#1}}
\newcommand{\Tab}[1]{Tab.~\ref{#1}}

\begin{document}

\title{
  Two-dimensional array of magnetic particles: 
  The role of an interaction cutoff 
}

\author{
  S. Fazekas$^{1,2}$, J. Kert\'esz$^{2}$, and D. E. Wolf$^{3}$
}

\affiliation{
  $^1$Department of Theoretical Physics,\\
  Budapest University of Technology and Economics,\\
  H-1111 Budapest, Hungary\\
  $^2$Theoretical Solid State Research Group
  of the Hungarian Academy of Sciences,\\
  Budapest University of Technology and Economics,\\
  H-1111 Budapest, Hungary\\
  $^3$Institute of Physics, University Duisburg-Essen,\\
  47048 Duisburg, Germany
}

\date{July 17, 2003}

\begin{abstract}
  Based on theoretical results and simulations, in
  two-dimensional arrangements of a dense dipolar particle
  system, there are two relevant local dipole arrangements: 
  (1) a ferromagnetic state with dipoles organized in a triangular
  lattice, and (2) an anti-ferromagnetic state with dipoles organized
  in a square lattice. In order to accelerate simulation algorithms 
  we search for the possibility of cutting off the interaction potential.
  Simulations on a dipolar 
  two-line system lead to the observation that the ferromagnetic state 
  is much more 
  sensitive to the interaction cutoff $R$ than the corresponding 
  anti-ferromagnetic state. For $R \gtrsim 8$ (measured in 
  particle diameters) there
  is no substantial change in the energetical balance of the 
  ferromagnetic
  and anti-ferromagnetic state and the ferromagnetic state slightly 
  dominates over the anti-ferromagnetic state, while the situation is 
  changed rapidly for lower interaction cutoff values, leading to the 
  disappearance of the ferromagnetic ground state. We studied the effect 
  of bending ferromagnetic and anti-ferromagnetic two-line systems and
  we observed that the cutoff has a major impact on the energetical
  balance of the ferromagnetic and anti-ferromagnetic state
  for $R \lesssim 4$. Based on our results we argue that 
  $R \approx 5$ is a reasonable choice for dipole-dipole interaction 
  cutoff in two-dimensional dipolar hard sphere systems, 
  if one is interested in local ordering.
\end{abstract}




\maketitle

\bigskip 

\section{Introduction} 

Long range interaction represents a major challenge for computer
simulations. The size of the tractable systems ($N$ particles) is
limited through the fact that the order of $N^2$ calculations are to
be carried out at each step, though for many purposes large systems
need to be studied. Periodic boundary conditions, which are
often helpful, can be implemented only by using sophisticated
summation algorithms (if possible due to screening).

In principle the above problems occur in so called short range
interaction models as well, like in the most extensively studied
Lennard-Jones system. However, for these systems a cutoff is usually
introduced making the original short range model explicitly finite
range. It is generally accepted that the error introduced by the cutoff 
is negligible provided the cutoff distance is large enough 
\cite{thij-comp-phys}.

Frequently the long range interaction potential falls off like $r^{-1}$
where $r$ is the distance between the particles. This should be
compared to the Lennard-Jones system where the potential decreases like
$r^{-6}$, where it is assumed that the attractive part of
that potential is due to induced dipole-dipole interaction.
 
In this study we focus on the question of cutting off a potential which
is in between the two above cases. We consider a two-dimensional
ensemble of magnetic particles interacting with a {$r^{-3}$} potential,
which, however, has an orientation dependence as well. It is crucial
from the point of view of efficient programing to know if a reasonable
cutoff can be introduced in this system. We investigate this problem by
comparing the stability of static configurations. 

\section{The Luttinger-Tisza method}

The Hamiltonian of a system of spherical dipoles is

\begin{equation}
  H = \frac{1}{2} \sum_{i \neq j} {\bf s}_i^T {\bf J}_{ij}~ {\bf s}_j,
  \label{eq:ham}  
\end{equation}

\noindent where $i$ and $j$ are dipole indexes, ${\bf s}$ denotes the 
dipole momentum vector, ${\bf s}^{T}$ denotes the transpose of 
${\bf s}$, and

\begin{equation}
  {\bf J}_{ij} = \frac{1}{\|{\bf r}_{ij}\|^3} 
  \left( 
    {\bf I} - 3~\frac{ {\bf r}_{ij} \circ {\bf r}_{ij} } 
      {\|{\bf r}_{ij}\|^2}
  \right),
  \label{eq:J}
\end{equation}

\noindent where ${\bf I}$ denotes the identity matrix, and 
${\bf r}_{ij}$ denotes the relative position vector of two dipoles. 
The $1/2$ factor in \Eq{eq:ham} avoids double counting of 
dipole pairs.

We can study the crystalline state of a dipole system using the 
Luttinger-Tisza method \cite{lut-tisz-pr46} based on the idea 
that in case of crystals it is a natural assumption that
the ground state exhibits some discrete translational symmetry.
If $\Gamma(i)$ denotes the points generated from $i$ with
discrete translations belonging to the $\Gamma$ symmetry group
the mentioned symmetry corresponds to ${\bf s}_i={\bf s}_{i'}$ for all
$i' \in \Gamma(i)$. According to this the system can be broken into
identical cells and the summation in \Eq{eq:ham} can be
limited to summation over one single cell. According to this 
the energy per dipole can be expressed as

\begin{equation}
  E = \frac{1}{2n}~ \sum_{i,j=1}^{n}
    {\bf s}_i^T {\bf A}_{ij}~ {\bf s}_j,
  \label{eq:erg-per-dip}
\end{equation}

\noindent where $n$ is the number of dipoles per cell, and 
${\bf A}_{ij}$ are symmetric matrices defined by

\begin{equation}
  {\bf A}_{ij} = 
    \sum_{ j' \in \Gamma(j),~ j' \neq i } {\bf J}_{ij'}.
  \label{eq:A-def}
\end{equation}

The expression of the energy per dipole in \Eq{eq:erg-per-dip} 
can be simplified by considering the 
$\hat{\bf s}=({\bf s}_i)_{i=1}^{n}$ 
hyper-vector and $\hat{\bf A}=({\bf A}_{ij})_{i,j=1}^{n}$
hyper-matrix. By construction $\hat{\bf A}$ is symmetric.
Using these, $E$ can be written as

\begin{equation}
  E = \frac{1}{2n}~ \hat{\bf s}^{T} \hat{\bf A}~ \hat{\bf s}.
\end{equation}

Solving the eigenvalue problem of the $nd$ dimensional 
symmetric matrix $\hat{\bf A}$, where $d$ is the dimension of the 
dipoles, we find the $\lambda_k$ eigenvalues and 
$\hat{\bf x}_k$ orthogonal eigenvector system 
with normalization
$\|\hat{\bf x}_k\|=\sqrt{n}$. 
Using these we have

\begin{equation}
  E = \frac{1}{2}~ \sum_{k=1}^{nd} \lambda_k b_k^2,
  \label{eq:eig-sum-erg-per-dip}
\end{equation}

\noindent where 
$b_k$ denotes the components of $\hat{\bf s}$ in the 
$\hat{\bf x}_k$ orthogonal eigenvector system. 

If the dipoles have identical scalar strength $\mu$,
i.e. $\|{\bf s}_i\|=\mu$, then $b_k$ must satisfy for all 
$i=1 \dots n$ the condition

\begin{equation}
  \left\| \sum_{k=1}^{nd} b_k {\bf x}_k^i \right\| = \mu,
  \label{eq:bk-strong}
\end{equation}

\noindent where ${\bf x}_k^i$ are the components of 
the $\hat{\bf x}_k$ hyper-vector which belong to dipole index $i$. 
Adding the square of the above equations and taking into consideration 
that $\{\hat{\bf x}_k\}_{k=1}^{nd}$ 
form an orthogonal system, we conclude that 
$b_k$ must satisfy the condition

\begin{equation}
  \sum_{k=1}^{nd} b_k^2 = \mu^2.
  \label{eq:bk-weak}
\end{equation}

In the framework of the Luttinger-Tisza method these two conditions are 
known as the strong (\Eq{eq:bk-strong}) and the 
weak (\Eq{eq:bk-weak}) conditions. From the weak condition 
and \Eq{eq:eig-sum-erg-per-dip} it can be derived that the 
energy per dipole in the ground state is 
$E_{min}=1/2~\lambda_{min}~ \mu^2$, 
where $\lambda_{min}$ denotes the smallest eigenvalue of $\hat{\bf A}$.
If there is one single eigenvalue equal to $\lambda_{min}$ the
ground state dipole arrangement is given by the corresponding
eigenvector. If there are more eigenvalues equal to $\lambda_{min}$, 
the ground state dipole arrangement is given by the linear combinations
of the corresponding eigenvectors which satisfy the strong 
condition.

\section{Two-dimensional array of magnetic particles}

The above method was applied to a system of two-di\-mensional 
dipole moments with identical scalar strength located at 
the sites of an infinite rhombic lattice with an arbitrary 
rhombicity angle by Brankov and Danchev \cite{bran-dan-pA87}.
They considered that the ground state of this system
has a translational symmetry
corresponding to discrete translations along
the lattice lines
with multiples of $2a$, where $a$ is the lattice constant. They
found that the ground state depends on the rhombicity
angle. 

We repeated their calculations with the consideration that the 
dipoles are carried by identical hard spherical particles of 
diameter equal to the lattice constant, and according to the 
geometrical constraint introduced by this consideration we limited the 
rhombicity angle to $60^{\circ} \leq \alpha \leq 90^{\circ}$. 
In accordance with their results we found that the system has a 
ferromagnetic ground state for 
$60^{\circ} \leq \alpha \lesssim 79.38^{\circ}$, 
and an anti-ferromagnetic
ground state for $79.38^{\circ} \lesssim \alpha \leq 90^{\circ}$.
We also found that the ground state for $\alpha=60^{\circ}$ is a 
continuously degenerate ferromagnetic state, and for
$\alpha=90^{\circ}$ is a continuously degenerate 
microvortex state including a four-fold degenerate anti-ferromagnetic 
state, where the microvortex state is defined as two anti-ferromagnetic 
sublattices making an arbitrary angle with each other. 
We also identified the states with the second lowest energy per dipole.
The results for $\alpha=60^{\circ}$ and $\alpha=90^{\circ}$ are
summarized in \Tab{tab:dip-cryst}.

\begin{table}[tbp]
\begin{center}
\begin{tabular}{|l|l|l|}
  \hline\hline  
  $\alpha=60^{\circ}$ 
    & $E = -2.758$ & continuously degenerate \\
    & & ferromagnetic state
  \\
  \cline{2-3} 
    & $E = -2.047$ & six-fold degenerate \\
    & & anti-ferromagnetic state
  \\
  \hline\hline
  $\alpha=90^{\circ}$ 
    & $E = -2.549$ & continuously degenerate \\
    & & microvortex state including \\
    & & a four-fold degenerate \\
    & & anti-ferromagnetic state
  \\
  \cline{2-3}
    & $E = -2.258$ & continuously degenerate \\
    & & ferromagnetic state
  \\
  \hline\hline
\end{tabular}
\end{center}
\caption{
  Dipole arrangements corresponding to the lowest two energy 
  values per dipole   
  in a two-dimensional system of dipole moments with identical 
  scalar strength located at 
  the sites of an infinite rhombic lattice with rhombicity angle
  $\alpha=60^{\circ}$ and $\alpha=90^{\circ}$. The energy is measured
  in $\mu^2/a^3$ units, where $\mu$ is the scalar strength of the
  dipole moments and $a$ is the lattice constant.
}
\label{tab:dip-cryst}
\end{table}

We repeated the calculations taking into consideration the 
interaction of only two neighboring lines on the rhombic lattice. This
corresponds to the interaction of two lines of 
dipolar hard spheres shifted according to the $\alpha$ rhombicity
angle. The Luttinger-Tisza method can be applied in a 
straightforward way also 
in this case. We observed that the ground state depends on 
the rhombicity angle $\alpha$ similar to the previous case. 
We found that the system has a ferromagnetic ground state for 
$60^{\circ} \leq \alpha \lesssim 75.67^{\circ}$,
and an anti-ferromagnetic ground state for 
$75.67^{\circ} \lesssim \alpha \leq 90^{\circ}$.
The ground state for $\alpha=60^{\circ}$ is a two-fold degenerate
ferromagnetic state, and for $\alpha=90^{\circ}$ is a two-fold 
degenerate anti-ferromagnetic state. 
We also identified the states with the second lowest energy per dipole,
and we summarized
the results for $\alpha=60^{\circ}$ and $\alpha=90^{\circ}$ 
in \Tab{tab:dip-line}.

\begin{table}[tbp]
\begin{center}
\begin{tabular}{|l|l|l|}
  \hline\hline  
    $\alpha=60^{\circ}$ 
      & $E = -2.582$ & two-fold degenerate \\
      & & ferromagnetic state
  \\
  \cline{2-3}
      & $E = -2.226$ & two-fold degenerate \\
      & & anti-ferromagnetic state
  \\
  \hline\hline
    $\alpha=90^{\circ}$ 
      & $E = -2.477$ & two-fold degenerate \\
      & & anti-ferromagnetic state
  \\
  \cline{2-3}
      & $E = -2.331$ & two-fold degenerate \\
      & & ferromagnetic state
  \\
  \hline\hline
\end{tabular}
\end{center}
\caption{
  Dipole arrangements corresponding to the lowest two 
  energy values per dipole   
  in a two-dimensional system of dipole moments with identical 
  scalar strength located on two neighboring lines 
  of an infinite rhombic lattice with rhombicity angle
  $\alpha=60^{\circ}$ and $\alpha=90^{\circ}$. The energy is measured
  in $\mu^2/a^3$ units, where $\mu$ is the scalar strength of the
  dipole moments and $a$ is the lattice constant.
}
\label{tab:dip-line}
\end{table}

It is not surprising that taking into consideration only two lines
of the rhombic lattice reduces significantly the original symmetry 
of the system. This can be seen comparing the results in 
\Tab{tab:dip-cryst} and \Tab{tab:dip-line}.
It is important to note that the two-line system has no continuously 
degenerate ground state, and thus the ground state is always defined by
the Luttinger-Tisza basic arrangement with the lowest eigenvalue.
In particular there is a special ferromagnetic and an 
anti-ferromagnetic state, which -- depending on the $\alpha$ rhombicity
angle -- define the ground state.

It can be also seen comparing the results in 
\Tab{tab:dip-cryst} and \Tab{tab:dip-line} that the 
interaction of two neighboring lines almost saturates the long-range 
dipole-dipole interaction, furthermore it is
widely known that dipolar spheres due to dipole-dipole interactions
tend to aggregate into chain like structures (see for example
\cite{weis-mp02,kun-prE99} and references therein) in which
the energies of intrachain interactions are much greater than those
of interchain interactions \cite{Rozen-prlB96}. These 
confirm that studying a 
two-line system gives valuable results related to properties
of dipole-dipole interaction in general.

\section{The role of an interaction cutoff}

Brankov and Danchev \cite{bran-dan-pA87} observed that 
the ground state of a system of dipoles on an infinite rhombic lattice
is sensitive to the dipole-dipole interaction range. An $R$ interaction 
cutoff distance can be introduced in a natural way with a slight 
modification of \Eq{eq:A-def} as 

\begin{equation}
  {\bf A}_{ij} = 
    \left. 
      \sum_{ j' \in \Gamma(j),~ j' \neq i } {\bf J}_{ij'} 
    ~\right|_{~r_{ij'}<R},
  \label{eq:A-def-cutoff}
\end{equation}

\noindent where $r_{ij'}$ denotes the distance of two dipoles, and 
the $r_{ij'}<R$ constraint represents the fact that 
the summation must only contain terms corresponding to pairs of
spherical dipoles closer to each other than the $R$ cutoff distance.
We measure the interaction cutoff value in $a$ units, where $a$ is 
the lattice constant equal to the particle diameter.

It can be seen from \Eq{eq:J} that the strength of the 
dipole-dipole interaction decays with $1/r_{ij'}^3$, and thus
the above expression for large $R$ can be arbitrarily close to the 
long-range limit in \Eq{eq:A-def}. 
This means that the numerical evaluation of Luttinger-Tisza 
states in general can be based on \Eq{eq:A-def-cutoff} 
if $R$ is big enough.

\begin{figure}[tbp]
\begin{center}
\begin{tabular}{r}
\includegraphics{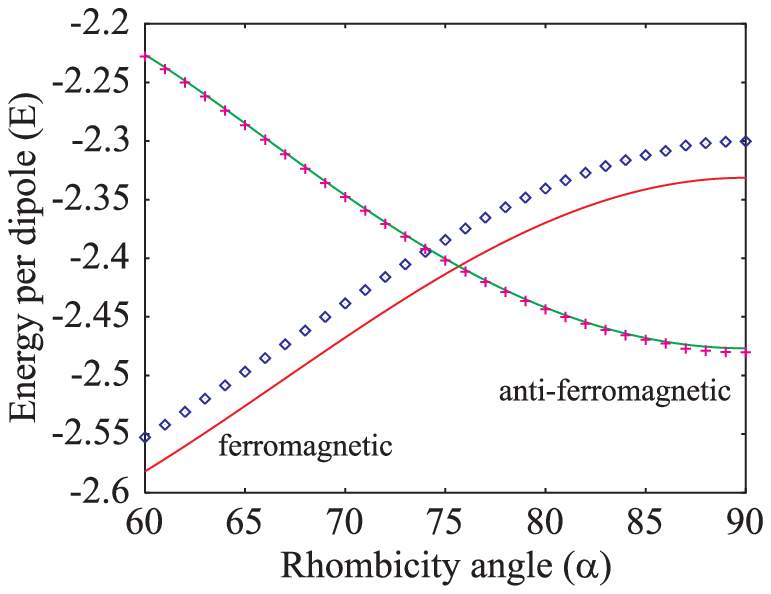} \\
\includegraphics{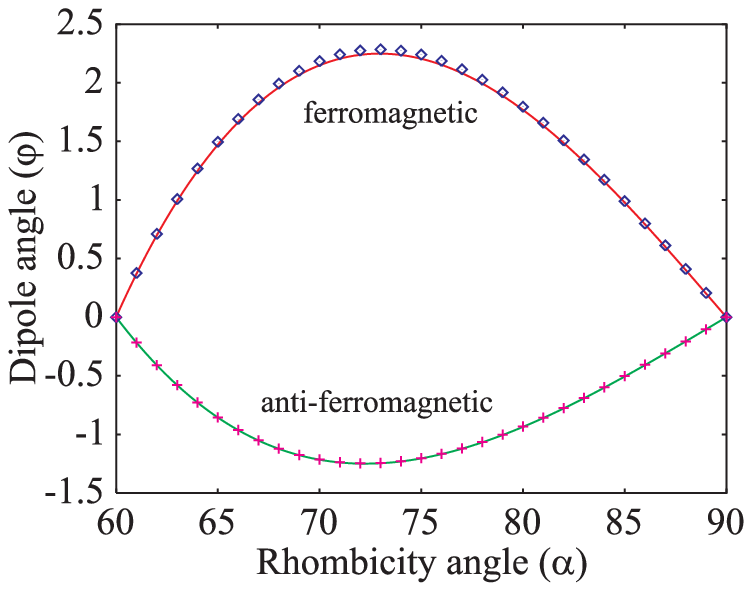} \\
\end{tabular}
\end{center}
\caption{
  Numerical results for the two-line system at $R$ equal to 
  $10^6$ (lines) and $8$ (points). The upper panel 
  shows the lowest energy per dipole of the ferromagnetic and the
  anti-ferromagnetic state 
  as function of the rhombicity angle. The energy is
  measured in units $\mu^2/a^3$. The lower panel shows the angle in
  degrees which the dipoles form with
  the direction of the longest linear dimension of the system.  
}
\label{fig:dip-line-R}
\end{figure}

Our numerical results for the two-line 
system at $R$ equal to $10^6$ (lines) and 
$8$ (points) are shown on \Fig{fig:dip-line-R}.
The results corresponding to $R=10^6$ are close to the
long-range interaction limit within the numerical errors of $64$ bit 
floating point arithmetic. The upper panel of \Fig{fig:dip-line-R}
shows the lowest energy per dipole of the ferromagnetic and the
anti-ferromagnetic state 
as function of the rhombicity angle. The lower panel of 
\Fig{fig:dip-line-R} shows the angle of the dipoles with
respect to
the direction of the longest linear dimension of the system.

Without any calculation one might expect that in the ground
state of the two-line system the dipoles are oriented parallel 
to the lines in both ferromagnetic and anti-ferromagnetic state.
It is a surprising result of our calculations that this is
true only for $\alpha=60^{\circ}$ and $\alpha=90^{\circ}$.
For any other $\alpha$ the dipoles form a small angle with the lines.
However these angles are less than $2.5^{\circ}$ they cannot be 
neglected.

Below a certain $\alpha$, as can be observed on 
\Fig{fig:dip-line-R}, the
ground state of the system corresponds to the ferromagnetic order, 
and above it to the anti-ferromagnetic order. 
The angle at which the transition from a ferromagnetic to an 
anti-ferromagnetic ground state happens, is shifted by only $3\%$
due to the cut off. That the anti-ferromagnetic state 
remains almost unchanged
is a consequence of the strong coupling of 
neighboring dipoles of opposite orientation, which makes 
the interaction cutoff irrelevant.

As expected (see \Fig{fig:dip-line-R})
the ferromagnetic state at $\alpha=60^{\circ}$ and the 
anti-ferromagnetic state at $\alpha=90^{\circ}$ are stable 
configurations of the two-line system of dipolar hard spheres
independent of $R$, however as $R$ decreases the crossover point 
gets slightly shifted.

\begin{figure}[tbp]
\begin{center}
\begin{tabular}{r}
\includegraphics{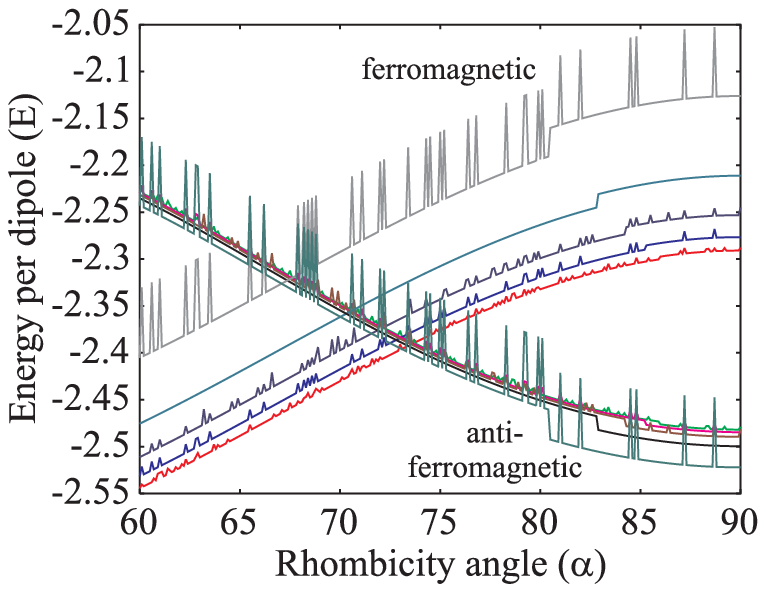} \\
\includegraphics{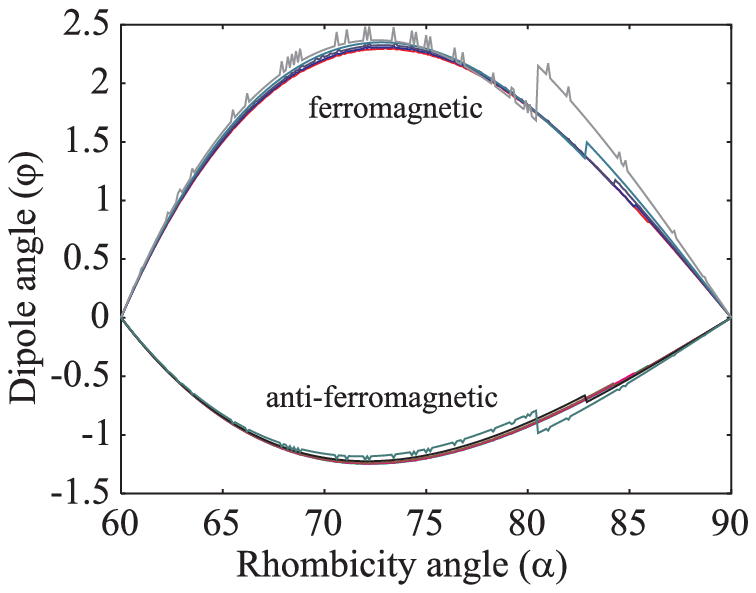} \\
\end{tabular}
\end{center}
\caption{
  Numerical results for the two-line system at $R$ equal to
  $7$, $6$, $5$, $4$, and $3$. The upper panel 
  shows the lowest energy per dipole of the ferromagnetic and the
  anti-ferromagnetic state 
  as function of the rhombicity angle. In ferromagnetic states 
  the lines are shifted upward as $R$ decreases.
  The lower panel shows the angle in degrees which the dipoles form with
  the direction of the longest linear dimension of the system.
}
\label{fig:dip-line-R2}
\end{figure}

At low interaction cutoff distances ($R \lesssim 8$) the discrete 
nature of the system becomes more and more 
relevant and both the ferromagnetic 
and anti-ferromagnetic energy per dipole begin to exhibit sudden jumps 
in function of the rhombicity angle (see \Fig{fig:dip-line-R2}).
As the interaction cutoff decreases the energy jumps become more
and more relevant.
This behavior can introduce numerical instabilities in simulations
using a badly chosen cutoff distance. It must be noted however that 
one should not overestimate this effect as the energy jumps are 
relatively small. It is an interesting observation that for some
interaction cutoff values (e.g. at $R \approx 4$) 
the energy per dipole shows significantly lower 
anomalies. 

The ferromagnetic line is shifted upward as $R$ decreases 
(see \Fig{fig:dip-line-R} and \Fig{fig:dip-line-R2}), and 
according to this the anti-ferromagnetic state becomes 
more and more dominant. For large $R$ the ferromagnetic state
at $\alpha=60^{\circ}$ has lower energy per dipole than
the anti-ferromagnetic state at $\alpha=90^{\circ}$.
Our numerical results show that at $R \approx 4$ the situation
is reversed, and at $R \approx 2$ the ferromagnetic ground state 
disappears. 
Brankov and Danchev \cite{bran-dan-pA87} found that
in case of an infinite rhombic lattice with 
rhombicity angle $\alpha=60^{\circ}$
the ferromagnetic ground state disappears at $R \approx 3$.

\section{Finite size corrections}

We investigated the finite size corrections of the energy per dipole
of the two-line system in ferromagnetic 
and anti-ferromagnetic states. In these states the
infinite system can be decomposed into identical finite segments.
If $N$ denotes the number of dipoles per line in a finite segment,
the energy per dipole of the infinite system can be written as

\begin{eqnarray}
  E = 
    \frac{1}{2N}~
    \left[~
      \frac{1}{2}~
        \sum_{ i,j \in \sigma(N), \; i \neq j } 
           {\bf s}_i^T {\bf J}_{ij}~ {\bf s}_j
    ~\right]
    + \nonumber\\
    \frac{1}{2N}~
    \left[~
      \frac{1}{2}~
        \sum_{ i \in \sigma(N), \; j \in \sigma(N)^c } 
          {\bf s}_i^T {\bf J}_{ij}~ {\bf s}_j
    ~\right]
    ,
\end{eqnarray}

\noindent where $\sigma(N)$ denotes the dipoles belonging to one
segment, and $\sigma(N)^c$ denotes the complementer of $\sigma(N)$.
The first part in the above expression can be recognized as the
energy ${\cal E}(N)$ per dipole of a finite segment containing 
$N$ dipoles per line. 

We define the following quantity
of energy dimension

\begin{equation}
  \partial{\cal E}(N) 
    \equiv 
     N~ \Big[~{\cal E}(N) - E ~\Big]
    .
\end{equation}

It can be seen that

\begin{equation}
  \partial{\cal E}(N) 
    = - \frac{1}{4}~
      \sum_{ i \in \sigma(N), \; j \in \sigma(N)^c } 
        {\bf s}_i^T {\bf J}_{ij}~ {\bf s}_j
    ,
\label{eq:F-def}
\end{equation}

\noindent where in case of an interaction cutoff $R$ one may add the 
condition $r_{ij}<R$.
As ${\bf J}_{ij}$ is proportional to $1/r_{ij}^3$ (see \Eq{eq:J}), 
one may expect that for large system size $\partial{\cal E}(N)$ is 
independent of $N$, and thus the limit

\begin{equation}
  \partial E = \lim_{N \rightarrow \infty} \partial{\cal E}(N),
\end{equation}

\noindent exists and is finite. Our numerical investigations confirmed
this expectation. The convergence of $\partial{\cal E}(N)$ 
is of order $1/N$ in the ferromagnetic case, and is of order 
$1/N^3$ in the anti-ferromagnetic case. 
This proves that the above quantity is well
defined. We refer to the above quantity as
the finite size coefficient. 

\begin{figure}[tbp]
\begin{center}
\begin{tabular}{r}
\includegraphics{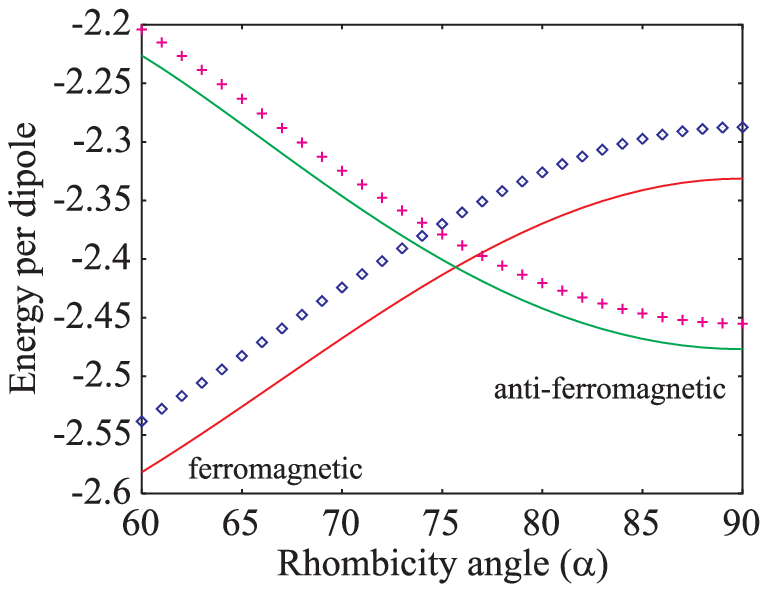} \\
\includegraphics{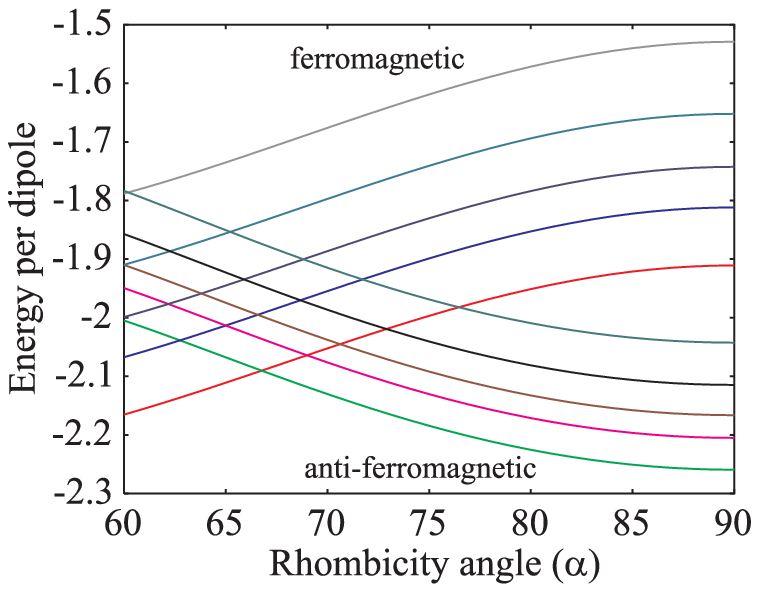} \\
\end{tabular}
\end{center}
\caption{
  Numerical results showing the dependence of the energy per dipole
  of the two-line system
  on the system size in the long-range limit. The upper panel
  shows results for $N$ (number of particles per line) equal to $10^5$ 
  (lines) and $100$ (points). The lower panel shows results for 
  $N$ equal to $10$, $8$, $7$, $6$, and $5$.
  Both the ferromagnetic and anti-ferromagnetic lines are moved upward 
  as $N$ decreases. (Note the different scales on the vertical 
  axes.)
}
\label{fig:fin-line-N}
\end{figure}

Numerical results showing the dependence of the 
energy per dipole ${\cal E}(N)$ of the two-line 
system on the system size 
in the long-range limit are presented in
\Fig{fig:fin-line-N}. The upper panel
shows results for $N$ equal to $10^5$ (lines) and $100$ (points), and 
the lower panel shows results for $N$ equal to $10$, $8$, $7$, 
$6$, and $5$. 
Both the ferromagnetic and anti-ferromagnetic lines are moved upward 
as $N$ decreases. ${\cal E}(N)$ at $N=10^5$ is close to the 
energy per dipole $E$ of the infinite two-line system 
within the numerical errors
of $64$ bit floating point arithmetic. 

Based on the definition of the finite size coefficient for large $N$ 
the energy per dipole of a finite system can be approximated as 

\begin{equation}
  {\cal E}(N) \approx E + \partial{E} / N.
  \label{eq:fin_erg_per_dip}
\end{equation}

Our numerical investigations show that this approximation
is reasonable even for $N \approx 10$. The finite size coefficient of
the ferromagnetic state is approximately two times bigger than the 
finite size coefficient of the anti-ferromagnetic state, and thus
the ferromagnetic line moves upward approximately two times faster
than the anti-ferromagnetic line (see \Fig{fig:fin-line-N}). 
It can be observed that for large $N$ the ferromagnetic state
at $\alpha=60^{\circ}$ has lower energy per dipole than
the anti-ferromagnetic state at $\alpha=90^{\circ}$.
Our numerical results show that at $N=20$ the situation
is reversed, and at $N=5$ the ferromagnetic ground state 
disappears. 

\section{Dependence of the finite size coefficient on interaction cutoff}

\begin{figure}[tbp]
\begin{center}
\begin{tabular}{r}
\includegraphics{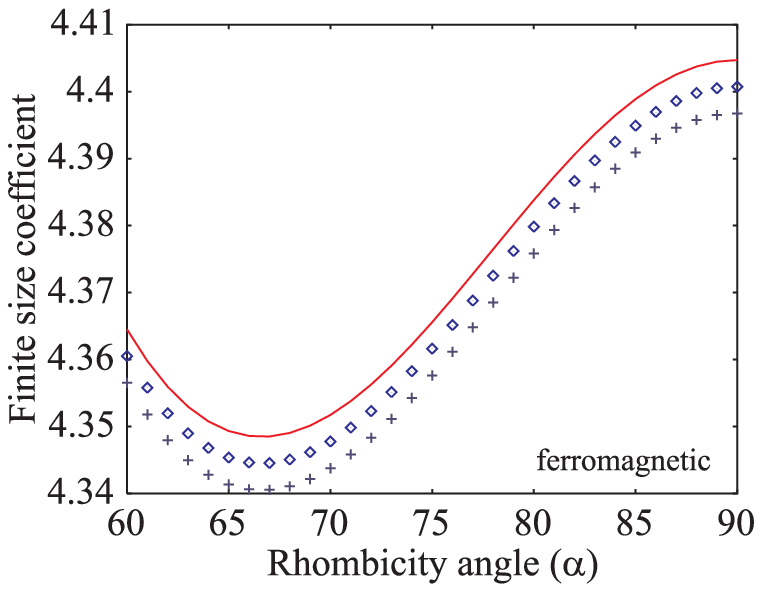} \\
\includegraphics{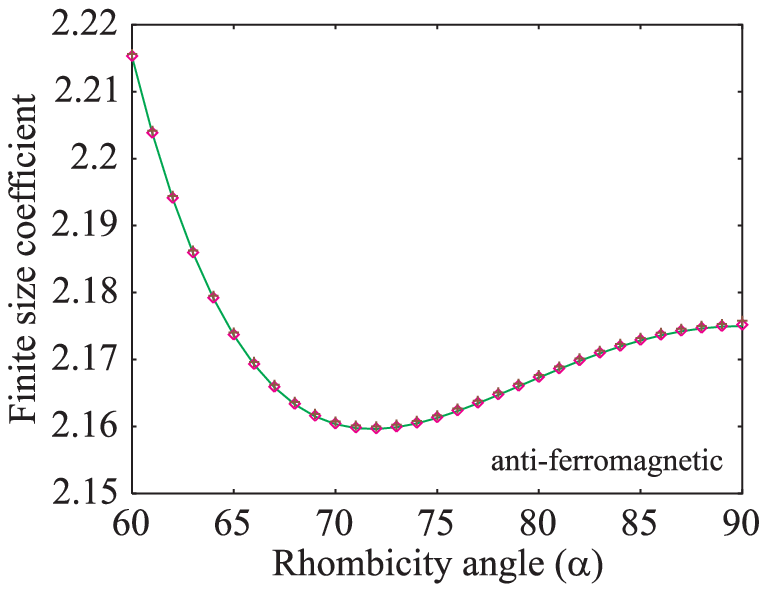} \\
\end{tabular}
\end{center}
\caption{  
  Numerical results showing the dependence of the finite size coefficient
  for both ferromagnetic and 
  anti-ferromagnetic state at $R=10^6$ (lines), for ferromagnetic state 
  at $R$ equal to $1000$ and $500$ (points), and 
  for anti-ferromagnetic state at $R$ equal to $100$ and $50$ (points). 
  The upper panel shows results 
  for the ferromagnetic state and the lower panel shows results for 
  the anti-ferromagnetic state. The finite size coefficient is 
  measured in units $\mu^2/a^3$. In the ferromagnetic case the lines
  are lowered as $R$ decreases. 
  (Note the different scales on the vertical axes.)
}
\label{fig:coef-fin-R}
\end{figure}

We investigated the dependence of the finite size coefficient 
on the interaction cutoff 
distance $R$. 
\Fig{fig:coef-fin-R} shows the numerical results for
ferromagnetic state (upper panel) at $R$ equal to $1000$ and $500$ 
(points), 
for anti-ferromagnetic state (lower panel) at $R$ equal to 
$100$ and $50$ (points),
and for both ferromagnetic (upper panel) and anti-ferromagnetic state 
(lower panel) at $R=10^6$ (lines). 
The finite size coefficient is measured in units $\mu^2/a^3$. 
We calculated its value 
by evaluating the expression in \Eq{eq:F-def} at 
$N=10^5$. The results for $R=10^6$ are close to
the long-range limit within the errors of $64$ bit floating point 
arithmetic. 

As $R$ is lowered in the anti-ferromagnetic case 
the finite size coefficient remains almost unchanged 
even for $R \approx 50$, while in the ferromagnetic case it
decreases significantly already at $R \approx 1000$. This 
shows again that the ferromagnetic state is much more sensitive to the
interaction cutoff than the anti-ferromagnetic state.

\begin{figure}[tbp]
\begin{center}
\begin{tabular}{r}
\includegraphics{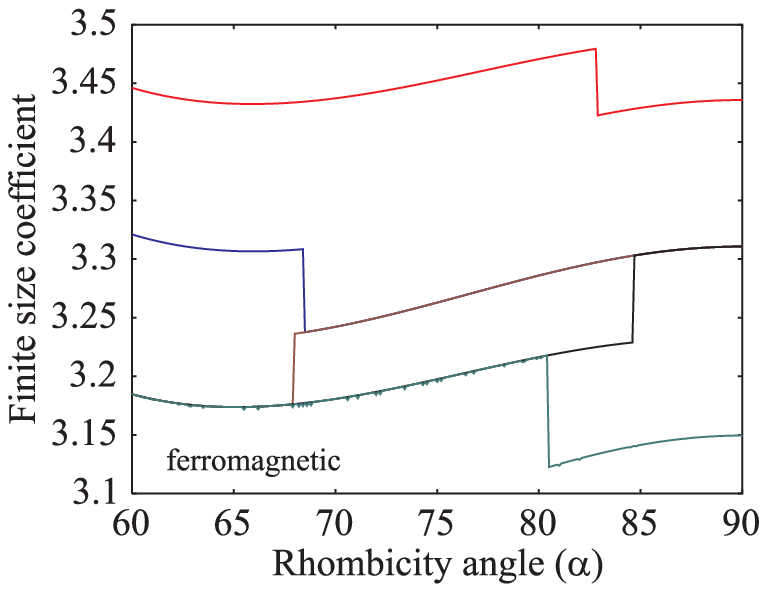} \\
\includegraphics{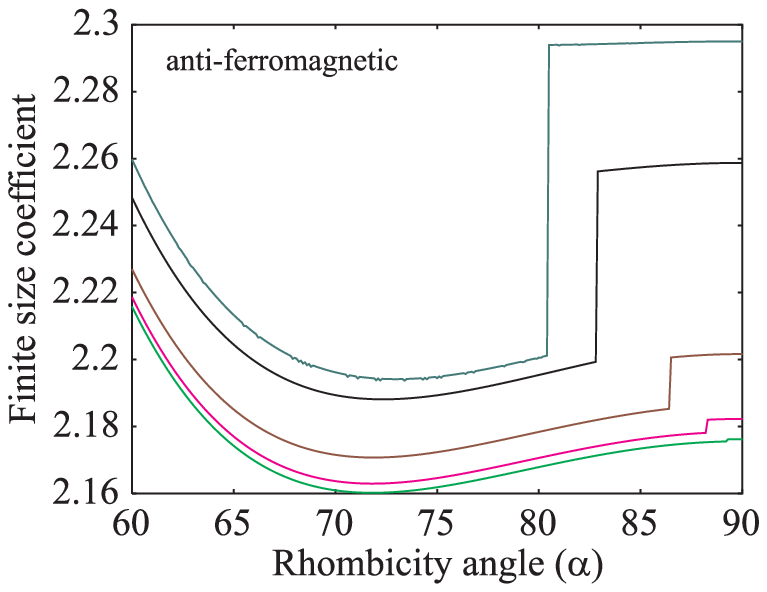} \\
\end{tabular}
\end{center}
\caption{
  Numerical results showing the dependence of the finite size coefficient
  at lower interaction cutoff distances. The upper panel shows the finite 
  size coefficient of the ferromagnetic state as function of the 
  rhombicity angle at $R$ equal to $4$, $3.75$, $3.5$, $3.25$, and $3$. 
  The lines are lowered as $R$ decreases.
  The lower 
  panel shows the the finite size coefficient of the anti-ferromagnetic 
  state at $R$ equal to $40$, $16$, $8$, $4$, and $3$.
  The lines are shifted upward as $R$ decreases. 
  (Note the different scales on the vertical axes.)
}
\label{fig:coef-fin-R2}
\end{figure}

At lower interaction cutoff distances (at $R \lesssim 50$)
the discrete nature of the system manifests itself in
sudden jumps in the finite size coefficient 
(see \Fig{fig:coef-fin-R2}). 

The upper panel of \Fig{fig:coef-fin-R2} shows the finite size 
coefficient of the ferromagnetic state as function of the rhombicity
angle at $R$ equal to $4$, $3.75$, $3.5$, $3.25$, and $3$. 
The lines are lowered as $R$ decreases.
The lower panel shows
the finite size coefficient of the anti-ferromagnetic state 
at $R$ equal to $40$, $16$, $8$, $4$, and $3$. 
The lines are shifted upward as $R$ decreases.

The jumps in the finite size coefficient 
become bigger as the interaction cutoff decreases 
(see \Fig{fig:coef-fin-R2}).
These jumps are not relevant at large $N$, but can introduce energy jumps
at lower dipole numbers, and thus can introduce local numerical 
instabilities in simulations, but this effect should not be overestimated
as the introduced energy jumps are relatively small.

\section{Bending two lines of magnetic particles}

\begin{figure}[tbp]
\begin{center}
\begin{tabular}{c}
  \begin{tabular}{c}
    \includegraphics{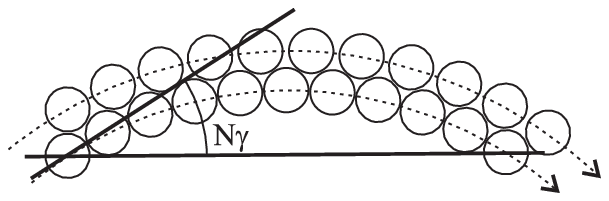}\\(a)\\~\\
    \includegraphics{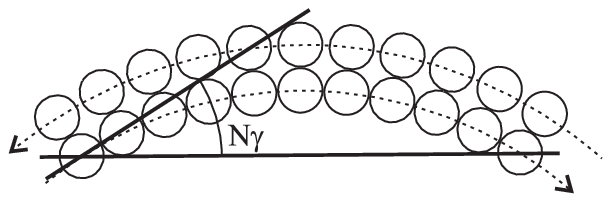}\\(b)
  \end{tabular} \\
  \begin{tabular}{c}
    \\\includegraphics{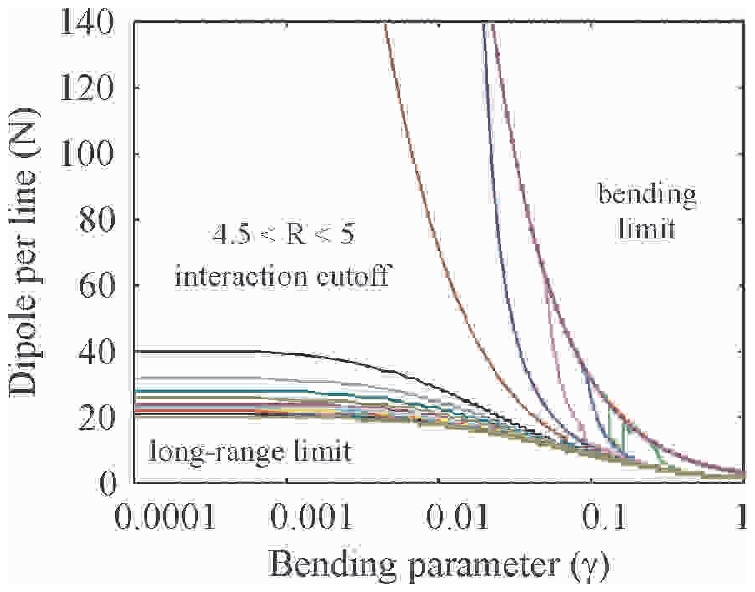}
  \end{tabular}
\end{tabular}
\end{center}
\caption{
  Numerical results related to bending a two-line system at different 
  interaction cutoff values. The upper panel shows a finite system of 
  two lines of dipolar hard spheres in ferromagnetic and 
  anti-ferromagnetic states. The lower panel shows $(\gamma, N)$
  state diagrams (see text for description) for $R$ ranging from
  $2$ to $\infty$. The lines are moved upward and lower as $R$
  decreases.
}
\label{fig:bent-sys}
\end{figure}

The finite size behavior presented before gives a good description
of finite dipole systems at large $N$, but it is not too helpful
at lower $N$. For a better understanding of the system,
we studied numerically finite systems (at small $N$) investigating 
the effect of bending two
lines of dipolar hard spheres in ferromagnetic and anti-ferromagnetic 
states (see \Fig{fig:bent-sys} (a) and (b)). 
In unbent case these correspond to the previously studied
ferromagnetic state at $\alpha=60^{\circ}$ and anti-ferromagnetic state 
at $\alpha=90^{\circ}$. We introduce the $\gamma$ bending parameter and 
we define the bent system as
composed of particles placed on an arc of angle $2N\gamma$ with dipole
vectors tangential to the arc (see definition of $\gamma$ on
\Fig{fig:bent-sys} (a) and (b)).
This definition involves a so called `bending limit' as the 
arc's angle is limited to $2\pi$, and thus $\gamma$ must satisfy 
the $\gamma \leq \pi/N$ condition. 

Our numerical results show that for bending either a ferromagnetic or
anti-ferromagnetic two-line system some physical effort is needed. We
observed that the two-line system in ferromagnetic state can be 
bent easier than in the corresponding anti-ferromagnetic state. This 
is a consequence of the strong coupling of neighboring dipoles
oriented anti-parallel. We also observed that as the anti-ferromagnetic 
state is bent it becomes less and less stable. \Fig{fig:bent-sys} 
shows numerical results related to bending a two-line system at different 
interaction cutoff values. As function of the bending 
parameter $\gamma$
and system size $N$ we compared the energy per dipole of 
the ferromagnetic and anti-ferromagnetic states and we identified the 
points $(\gamma, N)$ 
at which these two states are energetically equivalent. 
We repeated this procedure at different $R$ interaction cutoff values. 
The lower panel of \Fig{fig:bent-sys} shows corresponding 
$(\gamma, N)$ state diagrams. 

In the long-range limit for small system size and low 
bending parameter the anti-ferromagnetic state has lower energy per 
dipole. This is in accordance with our previous results, and
remains valid up to $R \approx 5$. It is a surprising result that
this behavior changes rapidly for interaction cutoff values between
$4$ and $5$. For $R \lesssim 4$ the anti-ferromagnetic 
state remains more stable at large $N$ values even for large bending 
parameters. This means that at this point 
the general characteristics of an arbitrary 
dipole system is substantially changed. Based on 
\Fig{fig:bent-sys} and on our previous results we argue that 
$R \approx 5$ is a reasonable choice for dipole-dipole
interaction cutoff for two dimensional systems of 
dipolar hard spheres, if one is interested in local ordering.

\section{Conclusions}

Based on the fact that
dipolar spheres due to dipole-dipole interactions
tend to aggregate into chain like structures in which
the ratio of interchain-to-intrachain interactions is small, 
and that moreover the interaction 
of parallel chains of dipolar hard spheres 
almost saturates the dipole-dipole
interaction in two-dimensional
dense systems, we argue that the study of
a dipolar two-line system gives valuable results for general 
dipolar particle systems. 

Theoretical results and simulations 
show two relevant dipole arrangements: 
(1) a ferromagnetic state with dipoles organized in a triangular
lattice, and (2) an anti-ferromagnetic state with dipoles organized
in a square lattice. Numerical results on a dipolar 
two-line system show that the ferromagnetic state is much more 
sensitive to the interaction cutoff than the corresponding 
anti-ferromagnetic state. This can be explained by the efficient coupling 
of dipoles oriented anti-parallel. For $R \gtrsim 8$ there
is no substantial change in the energetical balance of the ferromagnetic
and anti-ferromagnetic state and the ferromagnetic state slightly 
dominates over the anti-ferromagnetic state, while the situation is 
changed rapidly for lower interaction cutoff values, leading to the 
disappearance of the ferromagnetic ground state. 
Our numerical results show that 
the ferromagnetic ground state disappears at $R \approx 2$. 
Brankov and Danchev \cite{bran-dan-pA87} found that
in case of an infinite triangular lattice 
the ferromagnetic ground state disappears at $R \approx 3$.

For characterizing the finite size behavior of the two-line system
we introduced a finite size coefficient, and we observed
that it is sensitive to the interaction cutoff
for both ferromagnetic and
anti-ferromagnetic states. We also observed that at low interaction
cutoff values the discrete nature of the system leads to small
energetical anomalies. These anomalies increase as the interaction 
cutoff is lowered and can introduce instabilities in numerical
simulations. We argue however that this effect becomes relevant only
at first or second neighbor interaction and it can be neglected at 
higher interaction cutoff values. 

Finally we studied the effects of
bending ferromagnetic and anti-ferromagnetic two-line systems.
We characterized the bending of a two-line system with the
parameter $\gamma$, while $N$ is the number of dipoles per line.
We created $(\gamma, N)$ state diagrams separating energetically 
favorable ferromagnetic and anti-ferromagnetic states.
We observed that there is a substantial change of these state 
diagrams for $R \lesssim 4$, and -- in accordance with our previous
results -- we argue that $R \approx 5$ is a 
reasonable choice for dipole-dipole interaction cutoff in 
two-dimensional dipolar hard sphere systems, 
if one is interested in local ordering.

It is a surprising result that the reasonable interaction cutoff 
is independent of the strength of the dipole-dipole interaction and 
the particle size. This is a consequence of the fact that there
are two relevant dipole arrangements (a ferromagnetic and an 
anti-ferromagnetic), and their energetical balance can be reduced
to geometrical factors. If there are any other interactions in
the system (e.g. friction), this study must be revisited and it may
turn out that the reasonable interaction cutoff is dependent on
the interaction strength and particle size. We envision however
that in some cases (e.g. in case of friction) the presence of
another short-range interaction keeps or even lowers the value of 
the reasonable interaction cutoff found above.

In this paper we focused on the local dipole ordering.
In the ferromagnetic case however domain structures become important,
which can reduce external magnetic stray fields.
These global structures should depend on the long range part
of the interaction. For magnetic granular systems the formation
of such domains may be hindered e.g. by friction, though, as it 
requires the reorientation of particles.

We did not address the response to an external magnetic field.
The reason is that long range correlations and hence the response
functions will be more strongly affected by a dipolar interaction
cutoff than the local structures and energy densities considered in
this paper. In principle an Ewald summation method 
\cite{ewald-aph21,wang-holm-jchph01} would allow to 
explore the response properties in the thermodynamic limit in terms of
large but finite systems with periodic boundary conditions. However,
here again friction may be an important factor to be taken into
account: An external magnetic field trying to orient the magnetic
moments would exert a stress on the particle arrangement, if particle
rotations would be hindered by friction. Then the magnetic response of
the system would crucially depend on the relative strength of the 
magnetic anisotropy of the particles, i.e. the coupling between
particle and magnetic moment orientations, and friction forces between
the particles.

\section{Acknowledgments}


This research was carried out within the framework of the
``Center for Applied Mathematics and Computational Physics'' of the
BUTE, and it was supported by BMBF, grant HUN 02/011, and Hungarian 
Grant OTKA T035028.

\bigskip

\bibliography{magcutoff}
\bibliographystyle{apsrev}

\end{document}